\documentclass[12pt]{article}

\newcommand{\be}{\begin{equation}}
\newcommand{\dl}{\delta}
\newcommand{\lll}{\langle}
\newcommand{\rrr}{\rangle}
\newcommand{\ee}{\end{equation}}
\title{\bf
Perturbative Contributions to Field Correlators 
in Gluodynamics. }
\author{V.I.Shevchenko\thanks{e-mail: shevchenko@vxitep.itep.ru}
 and Yu.A.Simonov\thanks{e-mail: simonov@vxitep.itep.ru} \\
\it Institute for Theoretical and Experimental Physics \\
\it 117218, B.Cheremushkinskaya 25, Moscow, Russia}
\date{}
\begin{document}
\maketitle
\vspace{1cm}
{\centerline {\bf Abstract}}
The cancellation 
of perturbative contributions to the string tension
in gluodynamics in the framework of vacuum 
field correlators method
is shown at the order $O(g^4)$ by explicit calculation. 
The general pattern of these cancellations
at all orders and relation with the renormalization
properties of the Wilson loop is discussed.

\newpage

\section{Introduction}

Recently the formalism of gauge-invariant field correlators (FC) \cite{ss1}
has proved to be a useful tool in relating vacuum properties to the
QCD string parameters and hadron observables.
In particular the phenomena of confinement and deconfinement are
understood as due to particular terms of FC \cite{ss2} and the string
tension is obtained as an integral over those terms.

On the lattice FC have been measured both for $SU(2)$ \cite{di1}
and $SU(3)$ \cite{di2} gluodynamics and for the full QCD with the four
flavours of staggered fermions \cite{di3} and the first measurment
of FC in the vacuum without cooling was done recently \cite{di4}.
All these studies
refer to the nonperturbative contents of FC,
while the perturbative component, suppressed in the cooling process
on the lattice, is an admixture important at small distances
and seen clearly in \cite{di1}-\cite{di3}.
Analytically only the lowest order contribution $O(g^2)$ had been
known till recently, next-to-leading order terms
have been found in \cite{ej} and \cite{shev}. For the bilocal 
correlator (see (\ref{d})) the exact structure found in
\cite{ej,shev} looks like:
$$
\lll {\alpha}_s F(x) F(0) \rrr \sim \frac {a + b \ln\>x}{x^4}
$$
where $a$ and $b$ are constants.
The renormalization properties of FC (and the values of constants 
$a$ and $b$) cannot be 
entirely explained by charge 
renormalization and contain some additional contributions (see discussion 
in \cite{shev}).   
These results bring several questions, which are important for the
whole formalism of FC and which we try to answer below.

Firstly, what are renormalization properties of FC and how they are
connected with those of the Wilson loops? 
Secondly, the $O(g^4)$
contribution to the bilocal correlator $\lll{\alpha}_s F\>F\rrr$
formally leads to the (divergent) contribution to the
string tension, which physically has no sense and should be cancelled
by other terms. What is the exact mechanism of this cancellation?
And thirdly, one should see the general pattern of these cancellations
at all orders.

The paper is
organized as follows: Sect.2 is devoted to the definitions
of the essential ingredients of the formalism, in Sect.3
the exact relation between quadratic and triple correlators
is used to demonstrate the mechanism of cancellation of perturbative
contributions to the string tension at the order $O(g^4)$.
The Sect.4 concludes the paper with a discussion of the cancellation for higher
orders and the relation between  renormalization properties
of Wilson loops and FC.

\section{General definitions}

We start with 
the nonabelian Stokes theorem
\cite{stoks,cont} for the Wilson loop average
$W(C)$:  
$$
W(C) = \frac{1}{N_c}\>\lll Tr\>Pexp\>(ig\int\limits_{C} A_{\mu} dx^{\mu})\rrr
=
$$
\be
= \frac{1}{N_c}\>\lll Tr\>P_s exp\>(ig\int\limits_{S} F_{\mu\nu}(z,x_0)
 d{\sigma}_{\mu\nu}(z))\rrr
\label{www}
\ee
Here appears the basic quantity of the FC method - the field strength
operator $F_{\mu\nu}(z)$ covariantly  transported
with the help of the operators
$$
{\Phi}(z,x_0)=Pexp(ig \int\limits_{x_0}^{z} A_{\mu}(u) du_{\mu})
$$
to some chosen reference point $x_0$.
\be
F_{\mu\nu}(z, x_0) = {\Phi}(x_0, z) F_{{\mu}{\nu}}(z) {\Phi}(z, x_0)
\label{defin}
\ee
The averaging process, denoted in (\ref{www}) by angular brackets
is the standard integration in the QCD partition function, containing
gauge fixing and ghost terms. For our purposes we neglect quark degrees
of freedom and assume the perturbative expansion of the partition function,
yielding perturbative series for $W(C)$ and FC.

The cluster expansion theorem for (\ref{www}) reads:
\be
W(C) = \frac{1}{N_c}\> Tr\> exp\>\left( \sum\limits_{n=1}^{\infty}
\frac{(ig)^n}{n!} \int d\sigma(1)..d\sigma(n) \lll\lll F(1)..F(n)\rrr\rrr
\right)
\label{clus}
\ee
where we have suppressed the indices, $F(k) =
F_{\mu\nu}(z_k, x_0)$,
and we have used irreducible cumulants instead of averages, denoting them
with the double angular brackets \cite{kamp}. Note also that cumulants
are unit matrices in colour space and ordering operator in (\ref{clus})
is not needed in contrast to (\ref{www}).

Since $\lll\lll F(k)\rrr\rrr=0$, one can rewrite (\ref{clus})
identically as
\be
W(C) = \frac{1}{N_c}\> Tr\> exp\>\left( -\frac12
\int\limits_S \int\limits_S d{\sigma}_{\mu\nu}(u)d{\sigma}_{\rho\sigma}(v)
{\Lambda}_{\mu\nu , \rho\sigma}(u,v,C) \right)
\label{www1}
\ee
where we have defined the global correlator $\Lambda(u,v,C)$,
$$
{\Lambda}_{\mu\nu , \rho\sigma}(u,v,C) \equiv g^2 \lll\lll F_{\mu\nu}(u,x_0)
F_{\rho\sigma}(v,x_0)\rrr\rrr -
$$
\be
- 2\> \sum\limits_{n=3}^{\infty}
\frac{(ig)^n}{n!} \int d\sigma(3)..d\sigma(n) \biggl[ \lll\lll
F_{\mu\nu}(u,x_0)F_{\rho\sigma}(v,x_0)F(3)..F(n) \rrr\rrr +
\label{glob}
\ee
$$
+ \> perm.(1,2,..n)
\biggr]
$$
and $\>perm.(1,2,..n)$ stands for the sum of terms with different ordering of
$F(u,x_0)=F(1)$ and $F(v,x_0)=F(2)$ with respect to all other factors
$F(k)$. Since $W(C)$ does not depend on the shape of the surface $S$,
the dependence of the global correlator on its arguments is such,
that r.h.s. of (\ref{www1}) is independent of the choice of $S$ too
but depends on the contour $C$.
This circumstance explains the name "global correlator" used for
${\Lambda}(u,v,C)$ in contrast to local correlators which enter in the
r.h.s. of (\ref{glob}) (note however, that the
name "local" should not be misunderstood - correlators
$\lll\lll F(1)..F(k)\rrr\rrr$
depend on the points $z_1,..,z_k$ as well as
on the paths, entering in the definition (\ref{defin})
via transporters $\Phi(z,x_0)$).

Let us now fix one of the integration 
points in the exponent of (\ref{www1}) and denote the rest
integral as follows
\be
{Q}_{\mu\nu}(u,C) \equiv \frac12 \>\int\limits_{S} d{\sigma}_{\rho\sigma}
{\Lambda}_{\mu\nu , \rho\sigma}(u,v, C)
\label{www4}
\ee 
In the confining phase one expects for large contours 
$C$ the minimal area law of Wilson loop, which implies 
that $Q_{\mu\nu}$ in this limit does not depend on the point
$u$ when $S$ is the minimal area surface and simply coinsides with 
the string tension $\sigma$, while for the arbitrary surface 
one can identify $Q_{\mu\nu}$ as
\be
{Q}_{\mu\nu}(u,C) = P_{\mu\nu}\cdot\sigma
\ee
where $P_{\mu\nu}$ projects onto the minimal 
surface. Conversely if $Q_{\mu\nu}$ does not have constant limit
for large $S$ then the area law of Wilson loop does not hold.

To calculate $Q_{\mu\nu}(u,C)$ one can for simplicity 
take $x_0$ in (\ref{defin}) to coincide with $u$. Then the lowest
order FC in (\ref{glob}) depends only on two points (and on the straight
line, connecting them.) In what follows we concentrate on contributions 
to $\sigma$ and therefore take for simplicity a planar contour $C$
with the minimal surface $S$ lying in the plane.

The exact form of the two-point correlator may be written
in the following way \cite{ss2}:
$$
 D_{\mu\nu\rho\sigma}(u-v) =
Tr \lll gF_{\mu\nu}(u) \Phi(u,v) gF_{\rho\sigma}(v) \Phi(v,u)\rrr =
$$
\be
= \left( D(z^2) + D_1(z^2) + \frac{z^2}{2}\frac{dD_1(z^2)}{dz^2}\right)
{\Delta}^{(1)}_{\mu\nu\rho\sigma} - \frac{z^2}{2} \frac{d D_1(z^2)}{d z^2}\>
{\Delta}^{(2)}_{\mu\nu\rho\sigma} \label{d}
\ee
where two tensor structures
\be
{\dl}_{\mu\rho} {\dl}_{\nu\sigma} - {\dl}_{\nu\rho} {\dl}_{\mu\sigma} =
{\Delta}^{(1)}_{\mu\nu\rho\sigma} \ee
and
\be
{\Delta}^{(1)}_{\mu\nu\rho\sigma}
- 2 \left( \frac{z_{\mu}z_{\rho}}{z^2} {\dl}_{\nu\sigma} - \frac{z_{\nu}
z_{\rho}}{z^2} {\dl}_{\mu\sigma} + \frac{z_{\nu}z_{\sigma}}{z^2}
{\dl}_{\mu\rho} - \frac{z_{\mu}z_{\sigma}}{z^2}{\dl}_{\nu\rho} \right) =
 {\Delta}^{(2)}_{\mu\nu\rho\sigma}
\ee
were introduced.
Note, that
$$
{\Delta}^{(2)}_{\mu\nu\rho\sigma} {\dl}_{\mu\rho} {\dl}_{\nu\sigma} = 0
$$
therefore only the part proportional to ${\Delta}^{(1)}$ contributes
to the condensate $\lll {\alpha}_s F_{\mu\nu}F_{\mu\nu} \rrr $.

It was shown in \cite{ss2}, that the correlator $D_1(z)$ does not
contribute to the string tension, while the contribution 
of $D$ is
$$
{\sigma}^{(2)} = \frac12 \> \int d^2 v D(u-v)
$$
where the subscript $(2)$ refers to the quadratic correlator,
so that the total contribution of ${\Lambda}_{\mu\nu, \rho\sigma}$
can be written as the sum over contributions of correlators of order
$n$,
$$
\sigma = \sum\limits_{n=2}^{\infty} {\sigma}^{(n)}
$$ 
The perturbative studies of \cite{ej,shev} have revealed that
the lowest order contribution to $D(z)$ occurs at the $O(g^4)$
order. This implies that ${\sigma}^{(2)}$ is nonzero at this
order (actually it diverges), contrary to physical expectations.
It will be shown in the next section that there is another term 
at the same order of perturbatiom theory which exactly cancels
${\sigma}^{(2)}$. For that purpose one needs some relation between 
quadratic and triple correlators. The relation of this type, namely
the exterior derivative
of the function $D(z)$ from (\ref{d})
expressed through the triple correlators was found in \cite{ext}:
$$
 \varepsilon_{\mu_1\nu_1\sigma\rho}\frac{dD(z^2)}{dz^2}=\frac{i}{4}
 \varepsilon_{\mu_2\nu_2\xi\rho}
 \biggl(<Tr(F_{\mu_1\nu_1}(z_1)
 \tilde I_{\sigma\xi}(z_1,z_2)
 F_{\mu_2\nu_2}(z_2)\Phi(z_2,z_1))> -
$$
\be
 - <Tr(F_{\mu_1\nu_1}(z_1)
 \Phi(z_1,z_2) F_{\mu_2\nu_2}(z_2)
I_{\sigma\xi}(z_2,z_1)>\>\biggr)
\label{k1}
\ee
where
$$
  \tilde{I}_{\rho\gamma}(z,z')= \int^1_0 d\alpha\;\alpha\> \Phi
 (z,z+\alpha(z'-z)) F_{\rho\gamma}(z+\alpha(z'-z))\cdot
 $$
 \be
 \cdot \Phi(z+\alpha(z'-z),z')
 \ee
and analogously
 $$
 {I}_{\rho\gamma}(z,z')= \int^1_0 d\alpha\cdot\alpha \Phi
 (z,z'+\alpha(z-z')) F_{\rho\gamma}(z'+\alpha(z-z'))\cdot
 $$
 \be
 \cdot \Phi(z'+\alpha(z-z'),z')
 \ee
We have taken into accout the Bianchi identity
 $\varepsilon_{\mu_2\nu_2\xi\rho} D_{\xi} F_{\mu_2\nu_2}(z)=0$
 and denoted $z_2-z_1=z$. These relations will be important
in what follows.

In the framework of the described formalism it is natural to separate
perturbative and nonperturbative contributions to the functions $D(z)$
and $D_1(z)$ and take them into account differently for different
processes.
We are concentrating in the present paper
on perturbative parts of the bilocal and higher correlators to
explain several specific features the perturbation theory has in field
strength formulation.

\section{Cancellation of the perturbative contributions to the string tension
at the order $O(g^4)$. }

It has already been mentioned, that 
the results of \cite{ej,shev} imply that scalar functions $D(z)$ and
$D_{1}(z)$
both receive the perturbative contributions 
at the order $O(g^4)$ while at the tree level only
$D_1(z)$ is nonzero. The absense of perturbative contributions 
to the function $D(z)$ (and therefore to the string tension) at the 
tree level in $SU(2)$ gluodynamics was also noticed in different 
respect in \cite{deb}.

To look for cancellation at the given order 
$O(g^4)$ one must identify all terms of this order 
in ${\Lambda}(u,v,C)$ and $Q_{\mu\nu}(u,C)$. The $O(g^4)$
contribution comes from the quadratic and triple 
terms in (\ref{glob}) which we write in "polar" 
coordinates, $u=s_1z_1 ,\> v=s_2z_2, \> 0\le s_i \le 1$:
$$
{\Lambda}_{\nu\rho\mu\sigma} = g^2 \lll\lll F_{\nu\rho}(s_1z_1,x_0)
F_{\mu\sigma}(s_2z_2,x_0) \rrr\rrr +
$$
\be
+ \int\limits^{z_2} dz_3^{\phi} z_3^{\xi}\> \int\limits_0^1 ds_3 s_3
\lll\lll F_{\nu\rho}(s_1z_1,x_0) 
F_{\mu\sigma}(s_2z_2,x_0) F_{\xi\phi}(s_3z_3,x_0)
\rrr\rrr +
\ee
$$
+ O(\lll\lll FFFF\rrr\rrr)
$$
The term
$O(\lll\lll FFFF\rrr\rrr)$ starts from
the quartic cumulant and is $O(g^6)$, 
therefore it does not contribute to the function
$\Lambda_{\nu\rho\mu\sigma}$ at the $g^4$-order 
we are interested in at the moment.
 According to (\ref{www4}) we need to calculate
\be
{Q}_{\nu\rho} = \frac12 \int d V_{\beta}
{\epsilon}_{\mu\sigma\kappa\beta}\frac{\partial}{\partial v^{\kappa}}\>
{\Lambda}_{\nu\rho\mu\sigma}(u , v)
\ee
and show this quantity to be equal to zero at the desired order.
From the Stokes theorem point of view it means 
disappearance of the area term in the Wilson loop.

Since bilocal and triple cumulants coincide
with the usual correlators due to $\lll F_{\mu\nu}(z,x_0) \rrr = 0$
one gets (omitting for simplicity of notation 
the reference point $x_0 = 0$ and phase factors 
$\Phi(x_0, z)$ in all correlators):
$$
{\epsilon}_{\mu\sigma\kappa\beta}\frac{\partial}{\partial z_2^{\kappa}}\>
\biggl( \lll F_{\nu\rho}(s_1z_1)F_{\mu\sigma}(s_2z_2)\rrr +
$$
$$
+ \int\limits^{z_2} dz_3^{\phi} z_3^{\xi} \int\limits_0^1
ds_3 s_3 \lll F_{\nu\rho}(s_1z_1)F_{\mu\sigma}(s_2z_2)
F_{\xi\phi}(s_3z_3)\rrr\biggr) =
$$
\be
= {\epsilon}_{\mu\sigma\kappa\beta} \left( L_{DF} + L_3 + 
L_4\right)_{\kappa\nu\rho\mu\sigma}
\ee
where we have denoted
$$
L_{DF} =
\lll F_{\nu\rho}(s_1z_1)D_{\kappa}
F_{\mu\sigma}(s_2z_2) \rrr  $$

 $$ 
L_3 =
 \lll F_{\nu\rho}(s_1z_1)
 (s_2)^2 z_2^{\gamma} \int\limits_0^1
 d\alpha \alpha F_{\gamma\kappa}(\alpha s_2z_2) 
 F_{\mu\sigma}(s_2z_2) \rrr - $$ $$
- \lll F_{\nu\rho}(s_1z_1)
 (s_2)^2 z_2^{\gamma} \int\limits_0^1d\alpha \alpha 
 F_{\mu\sigma}(s_2z_2)
F_{\gamma\kappa}(\alpha s_2z_2)
 \rrr +
$$
$$
+ \int\limits_0^1 ds_3 s_3 z_2^{\gamma} \lll F_{\nu\rho}(s_1z_1)
F_{\mu\sigma}(s_2z_2) F_{\gamma\kappa}(s_3z_2) \rrr 
$$

$$
L_4 = \int\limits^{z_2} d z_3^{\phi} z_3^{\xi} \int\limits_0^1 ds_3 s_3
\biggl[\lll F_{\nu\rho}(s_1z_1) D_{\kappa} F_{\mu\sigma}(s_2z_2)
F_{\xi\phi}(s_3z_3) \rrr +
$$
$$
+  \lll F_{\nu\rho}(s_1z_1) (s_2)^2 z_2^{\gamma}
\int\limits_0^1d\alpha \alpha
 F_{\gamma\kappa}(\alpha s_2z_2)
 F_{\mu\sigma}(s_2z_2)
F_{\xi\phi}(s_3z_3) \rrr -
$$
\be
-  \lll F_{\nu\rho}(s_1z_1) (s_2)^2 z_2^{\gamma}
\int\limits_0^1d\alpha \alpha
 F_{\mu\sigma}(s_2z_2)
 \alpha s_2z_2) F_{\gamma\kappa}(\alpha s_2z_2)
 F_{\xi\phi}(s_3z_3) \rrr \biggr]
\label{gen}
\ee
The phase factors have been differentiated according to
\cite{mm}.
The terms $L_{DF}$ containing 
$D_{\mu} {\tilde F}_{\mu\sigma}$ vanish because of
 Bianchi identity.
The terms of the order $\lll F^3\rrr$ 
may be rewritten as (in the radial gauge
for simplicity, $\Phi(0, x) = 1$ ):
$$
{\epsilon}_{\mu\sigma\kappa\beta}\> z_2^{\gamma}
 \biggl(\int\limits_0^{s_2} u\>du \lll F_{\nu\rho}(s_1z_1)
[ F_{\gamma\kappa}(u z_2) F_{\mu\sigma}(s_2z_2) ] \rrr +
$$
\be
 \int\limits_0^{1} u\>du \lll F_{\nu\rho}(s_1z_1)
F_{\mu\sigma}(s_2z_2) F_{\gamma\kappa}(u z_2) \rrr
\label{kk}
\ee
At the $O(g^4)$-order one can easily observe antisymmetric 
color structure of the tree-point correlator: 
\be
\lll F^a(x) F^b(y) F^{c}(z) \rrr \propto f^{abc} D_{3}
\label{ff}
\ee
The above expression will be used to demonstrate 
vanishing of (\ref{kk}).

Taking into account the identity
$
f^{abc} t^a t^b t^c = {i}/{4} \left( N^2 - 1 \right) {\hat 1}
$
and performing the symmetrization with respect to integrations
over $s_2$ and $u$ one obtaines that the integrand in (\ref{kk}) is
proportional to
$$
 \lll F_{\nu\rho}(s_1z_1)
\{ F_{\mu\sigma}(s_2z_2) F_{\gamma\kappa}(u z_2) \} \rrr
$$
where $\{.. \}$ denote anticommutator.
This average is zero due to (\ref{ff}).

Situation with the $\lll F^4 \rrr $ terms in (\ref{gen}) 
is simpler, since only
disconnected parts of the quartic correlator contribute at the $g^4$ order.
Hence phase factors may be omitted and the correlator is
 factorized :
$$
\lll F_1 F_2 F_3 F_4 \rrr =
\lll F_1 F_2 \rrr \lll F_3 F_4\rrr
+ \lll F_1 F_3 \rrr \lll F_2 F_4\rrr
+ \lll F_1 F_4 \rrr \lll F_2 F_3\rrr
$$
It is easy to observe, that two last terms in (\ref{gen})
cancel each other at this order.
This finishes the proof of the stated cancellation at
 the $g^4$ order.

\section{Renormalization properties of the Wilson
loops and field correlators.}

The Wilson loop renormalization properties were studied 
in \cite{dotsenko,sato} and for smooth contour $C$ the result is:
\be W(C) = Z\cdot
W_{ren}(C) \label{ren}
\ee
where the (infinite) $Z$-factor contains linear
divergencies arising from the integrations over the contour
while all logarithmic divergencies are absorbed into
the renormalized charge  
$g_{ren}(\mu)$ defined at the corresponding dynamical scale $\mu$.

To connect the property (\ref{ren}) with FC one can write the perturbative series 
for $W(C)$ in the form of the cluster expansion:
$$
W(C) = \frac{1}{N_c}\>\lll Tr\>Pexp\>(ig\int\limits_{C} A_{\mu} dx^{\mu})\rrr=
$$
\be
= \frac{1}{N_c}\> Tr\> exp\>\left( -\frac12
\int\limits_C \int\limits_C d{z}_{\mu}d{u}_{\nu}
{\cal A}_{\mu\nu}(z,u,C) \right)
\label{www3}
\ee
where ${\cal A}_{\mu\nu}(z,u,C)$ is defined as (note an analogy with the definition 
of ${\Lambda}_{\mu\nu , \rho\sigma}$ in (\ref{glob}))
$$
{\cal A}_{\mu\nu}(z,u,C) \equiv g^2 \lll\lll A_{\mu}(z)
A_{\nu}(u)\rrr\rrr -
$$
\be
- 2\> \sum\limits_{n=3}^{\infty}
\frac{(ig)^n}{n!} \int dz(3)..dz(n) \biggl[ \lll\lll
A_{\mu}(z)A_{\nu}(u)A(3)..A(n) \rrr\rrr + perm.\biggr]
\label{globaa}
\ee
It is clear that $\lll A_{\mu}(z)\rrr = 0 $ and the ordering 
operator $P$ is not needed in (\ref{www3}) due to the color 
neutrality of the vacuum.

One can now use one of the coordinate gauges \cite{cont} and connect $A_{\mu}$ and
$F_{\mu\nu}$ , in the simplest Fock-Schwinger gauge one has:
$$
A_{\mu}(x) = \int\limits_{0}^{1} s\:x_{\nu} F_{\nu\mu}(s) ds
$$
As a consequence ${\cal A}_{\mu\nu}$ is expressed through FC as follows:
\be
{\cal A}_{\nu\sigma}(z,t,C) = \int\int  
\frac{ \partial u_{\mu}}{\partial z_{\nu}} 
\frac{ \partial v_{\rho}}{\partial t_{\sigma}} 
{\Lambda}_{\mu\phi , \rho\lambda}(u,v,C)
du_{\phi}
dv_{\lambda} 
\ee 
Now following the procedure of \cite{dotsenko,sato}
and comparing the perturbation series for ${\cal A}_{\mu\nu}$ and 
${\Lambda}_{\mu\nu , \rho\sigma}$ one can see that 
only those terms in ${\Lambda}_{\mu\nu , \rho\sigma}$
which take the form of the full derivatives in $u, v$ have 
counterparts in ${\cal A}_{\mu\nu}$ (i.e. the terms $D_{1}$ and
the similar structures for higher correlators) while the Kronecker 
type terms (proportional to ${\Delta}^{(1)}$) are cancelled since these
terms are not present in ${\cal A}_{\mu\nu}$. 
It is also clear, that  
all contributions arising from perturbative expansion of parallel
transporters in the gauge-covariant definition     
(\ref{defin}) are exactly cancelled at each given order 
$O(g^n)$ in ${\cal A}_{\mu\nu}$ (see also discussion of the 
related points in \cite{shev}). 
Hence perturbative contributions to the string tension are cancelled 
at any finite order $O(g^n)$ as well as those logarithmic 
contributions to the coupling constant renormalization which 
arise from the phase factors' perturbative expansion. 
Our explicit calculation in Section 
3 is the demonstration of this general statement in the special case 
$n=4$.

\bigskip

{\bf Acknowledgements}

The work was supported by RFFI grants 96-02-19184 and RFFI-DFG
96-02-00088G. The authors are grateful to H.G.Dosh for
stimulating discussions which initiated the present work and critical 
remarks.

\end{document}